\documentclass[12pt]{article}
\usepackage{amsmath}
\numberwithin{equation}{section}
\usepackage[left=1in,right=1in,bottom=1.75in]{geometry}
\usepackage{setspace}
\usepackage [linktocpage]{hyperref}
\date{(First dated: February 4, 2013)}
\begin{document}
\title{\bf New Instantons in AdS$_4$/CFT$_3$ from \\ D4-Branes Wrapping Some of CP$^3$ \ \\ \ }
\author{{\bf M. Naghdi \footnote{E-Mail: m.naghdi@mail.ilam.ac.ir} } \\
\textit{Department of Physics, Faculty of Basic Sciences}, \\
\textit{University of Ilam, Ilam, West of Iran.}}
%\date{\today}
 \setlength{\topmargin}{0.1in}
 \setlength{\textheight}{9.25in}
  \maketitle
  \vspace{0.2in}
    \thispagestyle{empty}
    \begin{center}
\textbf{Abstract}
\end{center}

With use of a 6-form field strength of ten-dimensional type IIA supergravity over $AdS_4 \times CP^3$, when $S^7/Z_k$ is considered as a $S^1$ Hopf fibration on $CP^3$, we earn a fully localized solution in the bulk of Euclideanized $AdS_4$. Indeed, this object appears in the external space because of wrapping a D4(M5)-brane over some parts of the respective internal spaces. Interestingly, this supersymmetry breaking $SU(4) \times U(1)$-singlet mode exists in already known spectra when one uses the $\textbf{8}_c$ gravitino representation of $SO(8)$. To adjust the boundary theory, we should swap the original $\textbf{8}_s$ and $\textbf{8}_c$ representations for supercharges and fermions in the Aharony-Bergman-Jafferis-Maldacena model. The procedure could later be interpreted as adding an anti-D4(M5)-brane to the prime $\mathcal{N}=6$ membrane theory resulting in a $\mathcal{N}=0$ antimembrane theory while other symmetries are preserved. Then, according to the well-known state-operator correspondence rules, we find a proper dual operator with the conformal dimension of $\Delta_+=3$ that matches to the bulk massless pseudoscalar state. After that, by making use of some fitting ansatzs for the used matter fields, we arrive at an exact boundary solution and comment on the other related issues as well.

\newpage
\setlength{\topmargin}{-0.7in}
\pagenumbering{arabic} % Roman page number
\setcounter{page}{2}   % Make it start with "ii"
                       %\tableofcontents

\section{Introduction}
Instantons and Solitons, as well-known nonperturbative effects, play many important roles in mathematics and physics especially. In the last few decades, their patterns in the gauge/string dualities have become even more important by the advent of AdS/CFT correspondence \cite{Maldacena}. The instantons have been widely studied for the famous duality of ten-dimensional (10d from now on) type IIB string theory over $AdS_5 \times S^5$ versus four-dimensional $\mathcal{N}=4$ $SU(N)$ Yang-Mills theory in \cite{Gibbons}, \cite{Green} and \cite{Bianchi1} firstly. Then, in mid 2008 and after releasing the best sample so far of AdS$_4$/CFT$_3$ duality by Aharony, Bergman, Jafferis and Maldacena (ABJM from now on) \cite{ABJM}, the instanton studies, next to many other efforts to discover various aspects of the model, got started first in \cite{Hosomichi}. Afterwards, through a few studies \cite{I.N}, \cite{N} and \cite{I}, we also found some new instanton solutions in the model.

Our first solution \cite{I.N}, which was in 11d supergravity, reduced the field equations over the skew-whiffed $AdS_4 \times S^7$ background to a conformally coupled scalar equation in the bulk of $AdS_4$. By detecting the exact solutions to the equation, we analyzed their behaviors near the boundary according to the well-known AdS/CFT correspondence rules \cite{Witten}. Later, to find the dual boundary operator, we exchanged the representations $\textbf{8}_s$ and $\textbf{8}_c$ of the membranes' boundary theory. The resulting theory was then for antimembranes. After that, by deforming the boundary theory by the founded operators, we arrived at some exact classical solutions in a sound correspondence with the bulk solutions.

The second solution \cite{N} was in 10d type IIA supergravity over the geometric background of $AdS_4 \times CP^3$. There, the localized solution in the bulk was a monopole instanton. In fact, we had a massless U(1) gauge field in the bulk of Euclideanized $AdS_4$ ($EAdS_4$) space whose excitation induced a magnetic field on the boundary. By turning on a boundary scalar field next to the $U(1) \times U(1)$ part of the full gauge group of $U(N) \times U(N)$ and making use of symmetries, we found the dual boundary operator and saw how the solutions of both sides matched clearly. This U(1) instanton was also studied in \cite{I} using a similar way except for an uplift of the exact bulk solution to the respective 11d supergravity.

In this paper, we continue the former lines of studies to find the instantons as the solutions with finite actions to the Euclidean equations of motion. It is remarkable that the new solution is more proper to be known as an equivalent for the famous D-instanton solution of AdS$_5$/CFT$_4$ duality studied in \cite{Green} and \cite{Bianchi1}. \\
We propose here an ansatz for the 5-form (6-form) field of the type IIA (M) supergravity version of ABJM while the main background geometry and fields are left unchanged. By doing so, we get a localized solution in the bulk of $EAdS_4$. The origin of the object in the bulk is likely from winding the added D4/M5-branes around some parts of the internal $CP^3$ or $S^7/Z_k$ of the complete 10d or 11d geometries. The ansatzs and solutions interestingly preserve the original symmetries but break all supersymmetries. It is indeed a pseudoscalar and a singlet of the isometry group of $SU(4) \times U(1)$ arisen from taking the prime $S^7$ as a U(1) Hopf fibration on $CP^3$. The basic motivation for the mode to be a pseudoscalar is its coming from the form fields with the internal space ingredients.\\
On the other hand, we see that in type IIA/M supergravity spectra of the involved Hopf fibration and Lens spaces (i.e. $AdS_4 \times S^7$, $AdS_4 \times S^7/Z_k$ and $AdS_4 \times CP^3$), as first traced in \cite{NilssonPope} and \cite{Sorokin1}, there is a singlet uncharged pseudoscalar in the bulk that matches to a marginal operator in a 3d $\mathcal{N}=0$ boundary CFT with the global symmetry of $SU(4)_R \times U(1)_b$. The last $U(1)\sim SO(2)$ becomes the baryonic symmetry in ABJM while the R-symmetry $SU(4) \sim SO(6)$ of the boundary field theory is the isometry group of $CP^3$. \\
Now, an important point is that the aforesaid pseudoscalar, which sits in the representation $\textbf{1}_0$ of $SU(4) \times U(1)$, exists just when the gravitinos (supersymmetry charges) are in the representations $\textbf{8}_c$ or $\textbf{8}_v$ of the original $SO(8)$ while the gravitinos are originally in $\textbf{8}_s$ of ABJM. So, in a similar line with \cite{I.N}, to adjust the bulk and boundary solutions, we should swap the representations $\textbf{c}$ and $\textbf{s}$ in ABJM. The resultant theory is then for antimembranes, and one may conclude that the branes, which we are wrapping over the internal spaces, are indeed anti-D/M-branes as we confirm more.

Next, to find a plain counterpart boundary solution, we first note that we have a massless pseudoscalar in the bulk and so, the dual boundary operator must have the conformal dimension of $\Delta_\pm=3,0$. The upper branch mode, which corresponds to the normalized bulk mode, is suitable. That is because the non-normalizable solutions are indeed not replying to the bulk fluctuations but they present some external sources which couple to the supergravity or string theory. Second, we note that the various terms in the $SU(4)_R \times U(1)_b$ -invariant Lagrangian of ABJM \cite{ABJM}, \cite{Terashima}, \cite{Klebanov} have the right dimension of 3. Third, it is proven that the deformations with the marginal boundary operators are often not deformations of the boundary theory but there are often new states in the same theory \cite{Balasubramanian}. Fourth, we may also look at the boundary operators for such bulk modes as proposed, for example, in \cite{Halyo1}, \cite{Bianchi2} and \cite{Forcella.Zaffaroni}.\\
All these facts suggest the operator with which we should deform the boundary theory. So, we handle an operator which has a similar structure as the Fermi's terms of the ABJM $SU(4)_R \times U(1)_b$-invariant Lagrangian. Another alternative operator, that one may use to adjust the bulk/boundary solutions, is the gauge parts of the mentioned Lagrangian similar to the process applied to find Yang-Mills instantons in $\mathcal{N}=4$ $SU(N)$ field theory--we should of course note that there are more subtleties with the Chern-Simons theories especially the types of deformations that one may employ. Afterwards, to match the bulk and boundary solutions, according to the gravity/gauge duality rules \cite{Witten}, we just turn on one scalar and one fermion alongside a $U(1)$ part of the full quiver gauge group of the model.

The Organization of this paper is as follows. In Sec. 2, we give a brief necessary review for the field theory and gravity side of the ABJM model. For the gravity side, we start from 11d supergravity and concisely arrive at 10d type IIA supergravity of the model. For the field theory side, we present the standard Lagrangian of the model alongside the needed symbols. In Sec. 3, we discuss the gravity side ansatzs, satisfy equations of motion, and inspect solutions, along with their associated interpretations and discussions. The spectra of the involved supergravities and how to arrive at our desired representation are also addressed. There, based on the founded solutions, we also evaluate the action and the added brane's charges, and briefly discuss the uplifting of the ansatz and solution to 11d supergravity. Section 4 is allocated to study and to find the dual field theory solutions and counterparts. There, we review the bulk-boundary correspondence rules for the case, set up the dual boundary operator and present a clear solution besides matching the bulk and boundary facts with a confirmation that the way is right. Section 5 includes summary, comments on supersymmetry and stability, and some other related issues and works to be addressed in future studies.

\section{A Brief of the Gravity/Gauge of the ABJM Model}
The ABJM \cite{ABJM} is so far the best known version for AdS$_4$/CFT$_3$ correspondence. It states that on the near horizon limit of a stack of N coincident M2-branes probing a singularity in $C^4/Z_k$ orbifold (which is indeed the IR limit), exists a three-dimensional $U(N)_k \times U(N)_{-k}$ Chern-Simons-matter theory at the level of $(k,-k)$ coupled to the matter fields in the bifundamental representation of the gauge group. The model has an $\mathcal{N}=6$ supersymmetry for generic $k$ that enhances to $\mathcal{N}=8$ nonperturbatively when the Chern levels are $k=1, 2$. For the last values of $k$, the model describes M2-branes in flat space and $R^8/Z_2$, respectively. The model is conjectured to have a dual gravitational description that is M-theory over $AdS_4 \times S^7/Z_k$ and, under some conditions, type IIA string theory over $AdS_4 \times CP^3$ as we describe more below.

\subsection{The Gravity Side of the Model}
To arrive at the near horizon limit of the model, one can start from the $AdS_4 \times S^7$ solution of 11d supergravity with $\acute{N}$$(=kN)$ units of the 4-form flux as follows:
\begin{equation}\label{eq01}
ds^2_{ABJM(M)}=\frac{R^2}{4} ds^2_{AdS_4}+R^2 ds^2_{S^7},
\end{equation}
\begin{equation} \label{eq02}
G^{(0)}_4 \approx \acute{N} \mathcal{E}_{AdS_4},
\end{equation}
where $R$, $\acute{N}$ and $\mathcal{E}_{AdS_4}$ are the curvature radius of 11d target-space, the initial number of flux quanta and unit volume-form of $AdS_4$, respectively. The $AdS_4$ metric in Poincar$\acute{e}$ upper-half plane coordinates, which we use here, with the Euclidean signature reads
\begin{equation}\label{eq03}
ds^2_{EAdS_4}=\frac{L^2}{u^2} \big(du^2+ dx_i dx_i \big), \quad i=1,2,3, %\sqrt{x^i x^i}=r %\quad \vec{u}=(x_1,x_2,x_3) \equiv r
\end{equation}
with a note that $2L=R=R_7=2R_{AdS}$. \\
One can always parametrize the transverse space to M2-branes through four complex coordinates $X_I$ ($I=1,2,3,4$) which are the needed coordinates (scalars) to embed the round seven-sphere $S^7$ as $\sum_{I=1}^4 |X_I|^2=1$. Now, by considering $S^7$ as an $S^1$ fibration on $CP^3$, one can write
\begin{equation} \label{eq04}
\begin{split}
ds_{S^7}^2 &=ds_{CP^3}^2+(d\acute{\varphi}+\omega)^2,
\end{split}
\end{equation}
where $\omega$ is a topologically nontrivial 1-form (that is dual to the Reeb killing vector of $\partial_{\acute{\varphi}}$) on $CP^3$, $\acute{\varphi}$ is the U(1) fiber coordinate with a period of $2\pi$, and the unit-radius metric of $CP^3$ with six specific real coordinates reads
\begin{equation}\label{eq04b}
 \begin{split}
ds_{CP^3}^2 &=d\xi^2+cos^2\xi sin^2\xi \bigg(d\psi+\frac{1}{2}cos\theta_1 d\varphi_1+\frac{1}{2}cos\theta_2 d\varphi_2 \bigg)^2 \\
& +\frac{1}{4} cos^2\xi \big(d\theta_1^2+sin^2\theta_1 d\varphi_1^2 \big)+\frac{1}{4} sin^2\xi \big(d\theta_2^2+sin^2\theta_2 d\varphi_2^2\big), \\
 \end{split}
\end{equation}
and
\begin{equation} \label{eq04c}
 \omega=\frac{1}{2} \big((cos^2\xi-sin^2\xi) d\psi+cos^2\xi cos\theta_1 d\varphi_1+ sin^2\xi cos\theta_2 d\varphi_2 \big),
\end{equation}
where $0 \leq \xi \leq \pi/2;\, 0 \leq \chi_s,\varphi_s, \acute{\varphi},\psi \leq 2\pi; \,  0 \leq \theta_s \leq \pi, \, s=1, 2$.\\
Here, the $Z_k$ quotient (orbifold) of $C^4$ acts on the four complex coordinates as $X_I \rightarrow e^{i2\pi/k} X_I$. Then, in order to have N units of the 4-from flux on the quotient space, one should take $\acute{N}=k N$ and $\acute{\varphi}=\varphi/k$ and so, the new metric reads
\begin{equation}\label{eq05}
   ds_{ABJM(IIA)}^2 = \tilde{R}^2 \big(ds_{AdS_4}^2+4ds_{CP^3}^2 \big), \quad \tilde{R}^2 =\frac{R^3}{4k}=\pi \sqrt{2\lambda},
\end{equation}
in which $\lambda\equiv N/k$ is the 't Hooft effective coupling constant of the boundary theory. In the interesting limit of large $N$ and for $\lambda \ll N^{1/5}$, the field theory is dual to M-theory over $AdS_4 \times S^7/Z_k$ together with $N$ units of $G^{(0)}_7$ flux on $S^7/Z_k$. When $k$ grows (the limit of $k\rightarrow\infty$ nearly), the M-theory circle shrinks and a better description for the dual field theory, in the limit of $N^{1/5}\ll k \ll N$, is type IIA string theory over $AdS_4 \times CP^3$ with $N$ units of the 6-form $F^{(0)}_6$ flux on $CP^3$ and $k$ units of the 2-form $F^{(0)}_2$ flux on $CP^1 \subset CP^3$. The form fields and dilation in type IIA theory are
\begin{equation}\label{eq06}
  e^{2\phi}=\frac{R^3}{k^3}, \quad H_3=dB_2=0, \quad F_2^{(0)}=dA_1^{(0)}=kJ, \quad F_4^{(0)}=dA_3^{(0)}=\frac{3}{8} R^3 \mathcal{E}_{4},
\end{equation}
where $\mathcal{E}_4$ is the $AdS_4$ unit volume-form and $J (=d\omega)$ is the K\"{a}hler form on $CP^3$.

\subsection{The Field Theory Side of the Model}
The three-dimensional $\mathcal{N}=6$ Chern-Simon-matter theory of ABJM is composed of the $U(N) \times U(N)$ gauge fields at the level of $k$ and $-k$ coupled to (anti) bifundamental matter fields. The theory can be constructed from theories with $\mathcal{N}=2$ and $\mathcal{N}=3$ supersymmetries, where two latter cases exist for any gauge group and charge content \cite{ABJM}. The $SU(4)_R \times U(1)_b$-invariant action of ABJM is always written as \cite{Terashima}, \cite{Klebanov}:
\begin{equation}\label{eq07}
 \begin{split}
   S_{ABJM}  =\int d^3x \,  \bigg\{\frac{k}{2\pi}\, \varepsilon^{\mu\nu\lambda} & \, tr \bigg(A_\mu A_\nu A_\lambda+\frac{2i}{3} A_\mu \partial_\nu A_\lambda-\hat{A}_\mu \hat{A}_\nu \hat{A}_\lambda-\frac{2i}{3} \hat{A}_\mu \partial_\nu \hat{A}_\lambda \bigg)\\
   & -tr\big(D_\mu Y_A^\dagger D^\mu Y^A \big)-tr\big(\psi^{A\dagger} i  \gamma^\mu D_\mu \psi_A \big)-V_{bos}-V_{ferm} \bigg\},
 \end{split}
\end{equation}
where the first pair of parentheses is for the Chern-Simons term while the second and third pairs are the kinetic terms for the bosons and fermions, respectively. The Bose scalar potential and Bose-Fermi interaction terms read
\begin{equation}\label{eq08}
   \begin{split}
   V_{bos} =-\frac{4\pi^2}{3k^2} tr&\big(Y^A Y_A^\dagger Y^B Y_B^\dagger Y^C Y_C^\dagger + Y_A^\dagger Y^A Y_B^\dagger Y^B Y_C^\dagger Y^C + 4 Y^A Y_B^\dagger Y^C Y_A^\dagger Y^B Y_C^\dagger \\
    & - 6 Y^A Y_B^\dagger Y^B Y_A^\dagger Y^C Y_C^\dagger\big),
  \end{split}
\end{equation}
 \begin{equation}\label{eq09}
   \begin{split}
   V_{ferm} =-\frac{2\pi i}{k} tr&\big(Y_A^\dagger Y^A \psi^{B\dagger}\psi_B - Y^A Y_A^\dagger \psi_B \psi^{B\dagger} + 2 Y^A Y_B^\dagger \psi_A \psi^{B\dagger} -2 Y_A^\dagger Y^B \psi^{A\dagger}\psi_B \\
   & +\varepsilon^{ABCD} Y_A^\dagger \psi_B Y_C^\dagger \psi_D- \varepsilon_{ABCD} Y^A \psi^{B\dagger} Y^C \psi^{D\dagger}\big),
   \end{split}
 \end{equation}
respectively. Here, $A_\mu, \hat{A}_\mu$ stand for $U(N) \times U(N)$ gauge fields. The matter fields, $Y^A$ and $\psi_A$ with ($A=1,..,4$), are four complex scalars and four three-dimensional spinor fields that each transforms in the bifundamental representation of the quiver gauge group as ($N,\bar{N}$). Besides the gauge symmetry, there is $SU(4)_R \times U(1)_b$ R-symmetry under which the scalars $Y^A$ transform as $\textbf{4}_1$ and the fermions $\psi_A$ transform as $\bar{\textbf{4}}_{-1}$. Meanwhile, the gauge covariant derivatives for the matter fields $\Phi$ ($Y^A$ or $\psi_{A}$) and the field strength of $F_{\mu\nu}$ read
\begin{equation}\label{eq10}
   \begin{split}
   & D_\mu \Phi =\partial_\mu \Phi+i A_\mu \Phi - i \Phi \hat{A}_\mu, \\
   & F_{\mu\nu}=\partial_\mu A_\nu-\partial_\nu A_\mu+i \big[A_\mu, A_\nu \big],
   \end{split}
 \end{equation}
respectively. The traces are taken on the gauge group $N \times N$ matrices meanwhile keeping the gauge-invariant quantities and setting the normalization of $U(N)$ as tr$(t^a t^b)=\frac{1}{2} \delta^{ab}$. The conventions for metric, Clifford's algebra and real gamma matrices in the original Minkowski signature read
\begin{equation}\label{eq11}
   \begin{split}
    \eta_{\mu\nu}=diag(-1,1,1), \ \ \{\gamma_\mu, \gamma_\nu \}=-2 \eta_{\mu\nu}, \ \ \gamma^\mu=(i\sigma_2,\sigma_1,\sigma_3), \ \ \varepsilon^{012}=1,
   \end{split}
 \end{equation}
where $\sigma_{1,2,3}$ are the usual Pauli matrices. We will see a small change of the relations in going to the Euclidean signature. Anyhow, various aspects of the Lagrangian and involved symmetries such as $\mathcal{N}=1, 2$ superfield formalism of the theory are studied in \cite{Gomiz}, \cite{Klebanov} among many others.

\section{New Instanton Solution in the Bulk of \textbf{\emph{AdS$_4$}}}
\subsection{The Ansatz in Ten Dimensions and Preliminaries}
We start with an ansatz for a 6-form field strength of type IIA supergravity, by making use of an established form in the ABJM model, as
\begin{equation}\label{eq12}
   \begin{split}
    & A_5=(f \ \omega \wedge J^2) \Rightarrow F_6=df \wedge \omega  \wedge J^2 + f \ J^3, \\
    & F_4= *_{10} F_6=*_4 df \wedge *_6(\omega  \wedge J^2) + k f (*_4 \textbf{1} \wedge *_6 J^3),
   \end{split}
 \end{equation}
where $f$ is a scalar function covering the whole $AdS_4$ space, $*_{10}\equiv *$ from now on, and we note that all coefficients are still included in the Hodge star.  \\
Now, by noting that the background geometry and fields in the model are kept unchanged; it is not difficult to check that all needed relations satisfy, interestingly. Clearly, the 10d type IIA supergravity action in string frame is given by
\begin{equation}\label{eq13}
\begin{split}
  S_{IIA} = & \frac{1}{2 \kappa^2} \int d^{10}x \, \sqrt{g} \, e^{-2\phi} \, R+\frac{1}{2 \kappa^2} \int \biggl\lbrack e^{-2\phi}  \, \bigl(4 d\phi \wedge \ast d\phi-\frac{1}{2} H_3 \wedge \ast H_3 \bigr) \\
  & -\frac{1}{2} F_2 \wedge \ast F_2-\frac{1}{2} \widetilde{F}_4 \wedge \ast \widetilde{F}_4 -\frac{1}{2} B_2 \, \wedge F_4  \, \wedge F_4 \biggr\rbrack,
\end{split}
\end{equation}
where $H_3=dB_2,\ F_2=dA_1,\ F_4=dA_3,\ \widetilde{F}_4=dA_3-A_1 \wedge H_3$ and the Hodge-star operation is taken with respect to the full 10d metric. By taking $H_3=0$, the same as ABJM, the relations to satisfy are
\begin{equation}\label{eq14a}
   dF_{p}=0,\quad d\ast F_{p}=0,
\end{equation}
\begin{equation}\label{eq14b}
   d\ast H_3 = g_s^2 (-F_2\wedge \ast \widetilde{F}_4+\frac{1}{2} \widetilde{F}_4 \wedge \widetilde{F}_4)=0,
\end{equation}
where $p=2,4$ and, in the last relation, use is made of the fact that the dilaton is constant with $e^{2\phi}=g_s^2$ for ABJM.\\
The arguments to satisfy the dilaton and metric equations are similar to those in our previous study \cite{N}. In fact, the dilaton equation is satisfied automatically while the rhs of the Einstein equations, on which the energy-momentum tensors are, remains to be satisfied. We, of course, may use the same reasonable tricks in \cite{N} to dissolve the problem. According to that, because the coefficient in front of the related energy-momentum tensors of the Einstein equations is $e^{2\phi}=R^3/k^3$, the added effect is negligible for the $k$ large enough to be the legality limit of the type IIA version of the ABJM model. Nevertheless, since the asymptotic symmetries of both sides of the duality remain unchanged, one may argue that the backreaction, if any, is tiny for our probe approximation especially. On the other hand, as long as we are interested in the behavior of the solutions near the boundary and correlation functions for dual operators, the backreactions on the background geometry can be ignored \cite{Bianchi3}, \cite{Skenderis}.

\subsection{Discussions on Solutions and Spectra}
One may proceed through the supersymmetry transformations for gravitinos in 10d or 11d to obtain the solutions and the number of preserved supersymmetries. However, here we go through satisfying the equations of motion directly.\\
The ansatz of (\ref{eq12}) satisfies $d*F_4=0$ trivially while to satisfy $dF_4=0$, the nontrivial conditions read
\begin{equation}\label{eq15}
     d (*_4 df)=0, \ \ \ d *_6(\omega  \wedge J^2)=0.
\end{equation}
By making use of (\ref{eq04b}) and (\ref{eq04c}), one can affirm that the second expression is satisfied, fortunately, whereas the first one, which is indeed the Laplace equation for $EAdS_4$, becomes
\begin{equation}\label{eq16a}
     \frac{1}{\sqrt{g}} \ \partial_{\acute{\mu}} \big(\sqrt{g} \ g^{\acute{\mu}\acute{\nu}} \partial_{\acute{\nu}} f \big)=\bigg[{\partial}_{i}{\partial}^{i}+u^2 \frac{\partial}{\partial u} \frac{1}{u^2} \frac{\partial}{\partial u}\bigg]f(u,\vec{u})\equiv L_4 f(u,\vec{u})=0,
\end{equation}
where $\acute{\mu}, \acute{\nu}, ...$ stand for four $AdS_4$ coordinates, and we define $\vec{u}= \vec{r}=(x_1,x_2,x_3)$. A familiar solution to this equation is
\begin{equation}\label{eq16bbcc}
      f(u,\vec{u};0,\vec{u}_0)=c_1 + \frac{c_2 \ u^3}{\big[u^2+(\vec{u}-\vec{u}_0)^2 \big]^3},
\end{equation}
where $c_1, c_2, ... $ are some constant coefficients related to the brane-instanton charges that we settle later. The solution is well-known, as it is also the Green function for a massless scalar propagating the instanton's location at $(0,\vec{u}_0)$ and another point at $(u,\vec{u})$. That is called the bulk to the bulk propagator and for the case that the source (instanton) is on the boundary of $AdS_4$, it is the boundary to the bulk propagator. Actually, the solution is singular at $u=0$ and corresponds to a small instanton on the boundary.

On the other hand, the field equation of (\ref{eq16a}) is for a massless scalar in $AdS_4$ and so, the conformal dimensions of the dual boundary operators, according to $\Delta_{\pm}=+\Delta \pm \sqrt{d^2 + 4 (m L)^2}/2$ for $AdS_{d+1}$, must be $\Delta_{\pm}=3,0$. Meanwhile, for the supergravity multiplets of the lowest mass, only the upper branch $\Delta_+$, which is the normalizable mode, is suitable. It is also notable that in the limit of approaching the boundary ($u\rightarrow 0$), the propagator of (\ref{eq16bbcc}) reduces to a delta function of $\delta^{(3)} (\vec{u}-\vec{u}_0)$. This singular point is the instanton position. In type IIB theory, the instanton was indeed a D(-1)-brane \cite{Green}, \cite{Bianchi1}, \cite{Chu}, \cite{Kogan}. What is that here? As we could see from the ansatz structure (\ref{eq12}), it may actually be interpreted as a kind of object coming from the Kaluza-Klein reduction of 11d or 10d supergravity on the related spaces in ABJM with the supplemented fields. In fact, it seems that, due to the wrapping of the world volume of the added Euclideanized electric D4-brane on the $CP^2 \times S^1$ part of the perfect internal space of $CP^3$, some fluctuations appear in the stature of the scalar of $f$ in the external space of $EAdS_4$.

A remarkable point toward the solution of (\ref{eq16bbcc}) is that it is a pointlike or a fully localized object in the 4d external space of $EAdS_4$. A counterpart to this in type IIB theory over $AdS_5 \times S^5$ is discussed in \cite{Chu} and \cite{Kogan} while the solution in \cite{Bianchi1} is localized in the whole 10d space. Therefore, the solution here smears on some parts of $CP^3$ or $S^7/Z_k$ likely to match the solutions on the boundary field theory. Whether the current solution can uplift to the 10d or 11d parent theories is related to the fact that the truncation is consistent or not.\footnote{A Kaluza-Klein truncation is consistent if only a finite set of the fields is maintained with a condition that the low-dimensional fields do not disturb the upper-dimensional ones or source them. Then, every solution to the low-dimensional theory is valid in the full upper-dimensional one.} We comment more on this point in the last section.

Another possible solution for the Laplace equation of (\ref{eq16a}) holds by separating the scalar function in its external variables as $f(u,\vec{u})=f(u) f(\vec{u})$. Then, in general, the $u$ part solution is a simple "exponential" function while the $\vec{u}$ part solution is a "distribution equation" in three dimensions. In the simplest case and after integrating out the three bulk coordinates of $\vec{u}$, which are indeed the D2(M2)-branes' world-volume directions, the smeared solution versus the localized solution reads $f(u)=c_3 + c_4 \ u^3$. In the latter case, the instanton is localized just in the $u$ direction and not in all Euclidean $AdS_4$ space.

An interesting point to say is that the operator $L_4$ in (\ref{eq16a}) is invariant under the conformal transformation of $ x_{\acute{\mu}} \leftrightarrow \frac{x_{\acute{\mu}}}{u^2+r^2}$ and so, the resultant solution goes to the last one and the order is reversed. This transformation maps a point at infinity to a point at the origin and exchanges the boundary conditions. However, the obstacle is that although the metric of (\ref{eq01}) or (\ref{eq05}) is also conformal invariant, the new form field of $F_6$, from which the solution (\ref{eq16bbcc}) arises, is not conformal invariant. So, more interesting discussions on the map are abandoned automatically.

Now, the question may be whether or not such a bulk excitation exists in the known spectra of the 11d and 10d supergravities over the involved spaces of $AdS_4 \times S^7/Z_k$ and $AdS_4 \times CP^3$. The answer is fortunately yes. Even so, we should first note that the object comes from a brane fully wrapping on the internal spaces (or a form field in terms of the known 1-form $\omega$ on the internal manifold) and so, it must be a pseudoscalar. Interestingly, there are pseudoscalar fluctuations in the gauged supergravities over such spaces \cite{NilssonPope}.\\
In fact, we note that there are three representations $\textbf{8}_s$, $\textbf{8}_c$, $\textbf{8}_v$ of the isometry group $SO(8)$ of $S^7$ for gravitinos. After the Hopf-fiber reduction, only the U(1)-neutral states remain in the spectrum. Moreover, we know that in the ABJM Lagrangian (\ref{eq07}), the supersymmetry charges (gravitinos), fermions and scalars decompose under the isometry group $SU(4) \times U(1)$ of $CP^3 \times S^1$ as
\begin{equation}\label{eq18a}
\begin{split}
  & \textbf{8}_s = \textbf{1}_{2} \oplus \textbf{1}_{-2} \oplus \textbf{6}_{0}, \\
  & \textbf{8}_c = \bar{\textbf{4}}_{-1} \oplus \textbf{4}_{1}, \\
  & \textbf{8}_v = \textbf{4}_{1} \oplus \bar{\textbf{4}}_{-1},
\end{split}
\end{equation}
respectively. This decomposition is indeed for the $\textbf{s}$ gravitino, for which the scalars and pseudoscalars are in $\textbf{35}_{v,c}$, while the gauge bosons are in $\textbf{28}$ for all cases. \\
The 35 scalars and 35 pseudoscalars, as well as gauge fields, from 11d gauged supergravity, decompose also as
\begin{equation}\label{eq18b}
\begin{split}
  & \textbf{35}_{v,c} = \textbf{10}_{2} \oplus \bar{\textbf{10}}_{-2} \oplus \textbf{15}_{0}, \\
  & \textbf{35}_s = \textbf{1}_{0} \oplus \bar{\textbf{1}}_{4} \oplus \textbf{1}_{-4} \oplus \bar{\textbf{6}}_{2} \oplus \textbf{6}_{-2}\oplus \bar{\textbf{20}}_{0},\\
  & \textbf{28} = \textbf{1}_{0} \oplus \bar{\textbf{6}}_{2} \oplus \textbf{6}_{-2} \oplus \textbf{15}_{0},
\end{split}
\end{equation}
respectively. For the $\textbf{c}$ gravitino, the only remaining scalars (pseudoscalars), in the massless spectrum of type IIA supergravity over $AdS_4 \times CP^3$, sit in $\textbf{15}_{0}$ ($\textbf{1}_{0}$); and for the $\textbf{v}$ gravitino, the only remaining scalars (pseudoscalars) sit in $\textbf{1}_{0}$ ($\textbf{15}_{0}$). Dual boundary theory is then a 3d $\mathcal{N}=0$ CFT theory with the global symmetry of $SU(4) \times U(1)$ and two marginal operators in $\textbf{1}_{0}$ and $\textbf{15}_{0}$. For the $\textbf{s}$ gravitino, only massless scalars (pseudoscalars) sit in $\textbf{15}_{0}$ and dual boundary theory is a 3d $\mathcal{N}=6$ CFT theory with the global symmetry of $SU(4) \times U(1)$ and two marginal operators just in $\textbf{15}_{0}$.

On the other hand, we note that the ansatz in (\ref{eq07}) is invariant (indeed singlet) under the full $SU(4) \times U(1)$ symmetry. It is because $\omega$ and therefore $J$ are invariant under $SU(4)$ and neutral with respect to $U(1)$. By knowing that we have a pseudoscalar ($0^-$) in the bulk, which exists in a skew-whiffed representation \cite{NilssonPope} (or "right representation" in the language of \cite{Duff}), fascinatingly, we are led to the statement that we are indeed winding (anti) branes around the internal spaces with right directions. When addressing the boundary side and state-operator correspondence, as well as in the last section, we return to the subject concisely.

\subsection{Charges and Actions}
Now, we try to evaluate the added (anti) brane charges. The electric charge of the included D4-brane based on the solution of (\ref{eq16bbcc}) with $c_1=0$, through the standard formula
\begin{equation}\label{eq19a}
 Q^{D4}_e=\frac{1}{\sqrt2 \kappa^2} \int \ast F_6,
\end{equation}
becomes
\begin{equation}\label{eq19b}
 Q^{D4}_e = c_5 \frac{k}{R^3}\frac{1}{\epsilon^3},
\end{equation}
where $\kappa^2=\frac{1}{2}(2\pi)^7$, $\epsilon > 0$ is a regulator small parameter \cite{Skenderis},\footnote{It should be mentioned that just when $x_{\acute{\mu}}=\epsilon \neq 0$, we have a definite charge.} and we have used the metric of (\ref{eq05}) and the following identities:
\begin{equation}\label{eq19c}
   \mathcal{E}_6 = \frac{1}{8.3!} J^3, \quad \ast J^3= \frac{k}{128 R^3} \ \mathcal{E}_4, \quad \ast \mathcal{E}_4= \frac{R^3}{3 k} \ J^3.
\end{equation}
One could also note that, to adjust to our symbol for $ F_2^{(0)}$ in (\ref{eq06}), we have taken the mentioned unit-volume element $\mathcal{E}_6$ for $CP^3$; therefore, we must in addition take $\int_{CP^1}J=2\pi$ for convenience. \\
One can similarly calculate the Euclideanized (anti-) D4-brane magnetic charge, which is indeed the electric charge of its dual (anti-) D2-brane. Now an interesting point about the charge in (\ref{eq19b}) is that, according to (\ref{eq05}), it is proportional to $\sim 1/{\sqrt \lambda}$. So, in the type IIA validity limit $\lambda\gg1$ of ABJM, it is almost negligible. This stresses our thought about ignoring the backreaction because of the added brane on the background.

Similar to the charges, we can estimate the corrections to the action because of the added fields. The relevant part of the main action of (\ref{eq13}) is now the fifth term. By inserting the ansatz of (\ref{eq12}), based on the solution of (\ref{eq16bbcc}) with $c_1=0$ into the action, we have
\begin{equation}\label{eq20a}
 \begin{split}
  S_{modi.}^{D4}= & - \frac{1}{256 (2\pi)^4}\frac{k^3}{R^3} \int_{AdS_4} f^2 dVol(AdS_4) \\
  & - \frac{1}{2 (2\pi)^7} \int_{AdS_4} (df \wedge *_4 df) \ \int_{CP^3} (\omega  \wedge J^2) \wedge *_6(\omega  \wedge J^2),
 \end{split}
\end{equation}
where $dVol(AdS_4)= \mathcal{E}_4$. The singular points of the integrals are at $u=0$ and so according to the regularization discussions \cite{Bianchi3}, \cite{Skenderis}, we may again keep just the finite part of the action. That is
\begin{equation}\label{eq20b}
 S_{modif.}^{D4} = c_6 \frac{k}{R^3}\frac{1}{\epsilon^6},
\end{equation}
on the boundary at $u=\epsilon$. We see again that the correction is a small amount.

\subsection{On the Ansatz Uplift to Eleven Dimensions }
The gravitational field spectra, which are chiral primaries on the ABJM background, are actually the projections of the primary spectra over $AdS_4 \times S^7$ into $Z_k$-invariant states \cite{ABJM}. Positively, the orbifold of $Z_k$ preserves the $SU(4) \times U(1)$ symmetry of the full $SO(8)$ isometry symmetry of $S^7$, as the various decompositions under $SO(8)\rightarrow SU(4) \times U(1)$ are given in (\ref{eq18a}). For $ k \geq 3$, two single supercharges in $\textbf{8}_s$ of the original theory are projected out and the remaining symmetry is just $\mathcal{N}=6$. For $k=1,2$, the supersymmetry enhances to $\mathcal{N}=8$ because of the "monopole operators" nonperturbatively.

In the lens-space of $S^7/Z_k$ and for $ k \geq 3$, the pattern is almost identical with that of $CP^3$. Indeed, for the skew-whiffed cases (the gravitinos in $\textbf{8}_{v,c}$), the boundary theory is a 3d $SU(4) \times U(1)$ $\mathcal{N}=0$ CFT theory with two marginal operators for the massless scalars (pseudoscalars) in $\textbf{1}_{0}$, $\textbf{15}_{0}$. For the gravitino of $\textbf{8}_{s}$, there is a 3d $SU(4) \times U(1)$ $\mathcal{N}=6$ SCFT theory with two marginal operators for the massless scalars (pseudoscalars) in $\textbf{15}_{0}$ \cite{NilssonPope}, \cite{Halyo2}.\\
Therefore, because our ansatz obviously breaks supersymmetry and that, at least for $k=1$, the skew-whiffed solution with $S^7$ is supersymmetric, we infer that the ansatz may not be applicable to the case. For $k=2$, as well, there is the maximal supersymmetry of $\mathcal{N}=8$ in the bulk for all gravitinos. Thus, it also excludes because our solution makes differences between $\textbf{8}_{v,c}$ and $\textbf{8}_{s}$ and breaks all supersymmetries.

Anyway, the main question here is the uplifting of the 10-dimensional ansatz of (\ref{eq12}) to an 11-dimensional ansatz. The best consistent ansatz may be
\begin{equation}\label{eq21a}
   F_7=df \wedge d\varphi \wedge \omega  \wedge J^2 - f \ d\varphi \wedge J^3.
 \end{equation}
Inserting the ansatz into the 11-dimensional form-field identity and equation,
\begin{equation}\label{eq22}
   dF_4=0, \qquad d*_{11} F_4 + \frac{1}{2} F_4 \wedge F_4=0,
 \end{equation}
we see that the equation is satisfied trivially while the identity is satisfied with the same Laplace equation of (\ref{eq16a}) and also if
 \begin{equation}\label{eq21b}
     d *_7 (d\varphi \wedge \omega  \wedge J^2)=0,
\end{equation}
which is not satisfied of course. Nevertheless, it does not appear to cause any serious problem because the original 7-form, which couples to the electric M5-branes, is satisfied and now its magnetic dual may be a partial solution and not a complete solution. Again, the ansatz and solution are $SU(4) \times U(1)$- invariant and so, along with other discussions in the subsection, one may follow the lines in the type IIA case. \\
By the way, we should remind the reader again that, for $k=1,2$, we do not have the founded mode in the known 11d supergravity over $AdS_4 \times S^7/Z_k$. On the field theory side, the associated chiral operators are $SO(8)_R$-invariant while the gravity solution here is $SO(6)_R$-invariant. So, these two special cases are not mainly included in our discussions. \\
It is also notable that one could place $e^7_{{S^1}/{Z_k}}={{\frac{1}{k}}}\left(d\varphi +k\omega \right)$ instead of $d\varphi$ into the ansatz of (\ref{eq21a}). Now, the added M5-brane probably wraps around $S^6/Z_k \sim CP^2 \times S^1 \times S^1$.

\section{Dual Boundary Solutions and Correspondence}
\subsection{Matching Bulk to Boundary}
Here, we explore a dual description for the bulk pseudoscalar mode based on the AdS/CFT correspondence prescriptions \cite{Witten}, \cite{KlebanovWitten}. Following the discussions on the spectrum, we first note that the ansatz of (\ref{eq12}) is a singlet of $SU(4) \times U(1)$. That is because $J$ and therefore $\omega$ and $e^7_{{S^1}/{Z_k}}\equiv e_7$ are $SU(4)$-invariant, and $J$ and $e_7$ do not carry any $U(1)$ charge. So, the dual boundary operator must be a singlet of $SU(4)_R \times U(1)_b$; see also \cite{I.N}. \\
The next question is which dual boundary operator is associated with the bulk sate. First, we note that turning on the normalizable mode is always considered as a different state in the same theory and not necessarily as a deformation of the original theory \cite{Balasubramanian}. This fact, next to some operators dual to such bulk states proposed for instance in \cite{Halyo1} and \cite{Bianchi2}, makes our task easier.

On the other hand, for a scalar field in the Euclidean $AdS_4$ and Poincar$\acute{e}$ upper half-plane coordinates, the asymptotic behavior of the solution near the boundary at $u=0$ is
\begin{equation}\label{eq23}
    f(u,\vec{u}) \approx u^{\Delta_-} \alpha(\vec{u}) + u^{\Delta_+} \beta(\vec{u}),
\end{equation}
where $\Delta_{\mp}=0, 3$ are for the solution of (\ref{eq16a}). $\alpha$ and $\beta$ have a holographic interpretation as "source" (the boundary value of the bulk field) and "one-point function" for the operator with the conformal dimension of $\Delta_+$, respectively and conversely for the $\Delta_-$ operator. Such a scalar can be quantized by either Dirichlet boundary condition $\delta\alpha=0$ (which can be used for any $m^2$) or Neumann boundary condition  $\delta\beta=0$ (which can be used when the scalar masses are in the range of $-9/4<m^2 L^2<-5/4$). In the "usual" CFT \cite{KlebanovWitten}, the $\alpha$, as source, couples to an operator with $\Delta_+$ (the normalizable mode).

Now, for the normalizable mode ($\Delta_+=3$) of the massless pseudoscalar, with the solution of (\ref{eq16bbcc}) at hand, we can write
\begin{equation}\label{eq24a}
    \alpha(\vec{u})=f_0(\vec{u}), \qquad \beta(\vec{u})=\frac{c}{|\vec{u}-\vec{u}_0 |^6} \equiv \frac{c}{r^6},
\end{equation}
where $c_2=c$, and we note that the first term in (\ref{eq23}) dominates as $u\rightarrow0$. Then, with the localized source of $f_0(\vec{u}_0)=\delta^3(\vec{u}-\vec{u}_0)$, we have
\begin{equation}\label{eq24c}
    \beta(\vec{u})=\frac{1}{3} \langle \mathcal{O}_3(\vec{u})\rangle_\alpha=-\frac{\delta W[\alpha]}{\delta\alpha}= \frac{\delta S_{on-shell}}{\delta \alpha(\vec{u})},
\end{equation}
where $\mathcal{O}_3$ stands for the boundary operator of $\Delta_+=3$, $W$ is the field theory "generating functional" and $S_{on-shell}$ is the bulk "on-shell" action. This means that, with the pseudoscalar bulk mode turned on, one should correct the boundary action as $S \rightarrow S + W$ \cite{Witten2} with
\begin{equation}\label{eq24d}
     W= - \frac{1}{3} \int d^3 \vec{u} \ \alpha(\vec{u})\ \mathcal{O}_3(\vec{u}),
\end{equation}
and we should also note that here $\alpha=c_1$, which we set to 1.

\subsection{The Boundary Solution}
According to the arguments already mentioned for the dual boundary operator $\mathcal{O}_3$ and that it may have the same structure as the ABJM Lagrangian's terms, as well as the proposed operators in \cite{Halyo1}, \cite{Bianchi2}, \cite{Forcella.Zaffaroni}, we employ the following operator:
\begin{equation}\label{eq25}
     \mathcal{O}_3= tr \big(Y_A^\dagger Y^A \psi^{B\dagger}\psi_B \big),
\end{equation}
where the matter fields transform in the same representation of the $SU(4)_R \times U(1)_b$-invariant Lagrangian of (\ref{eq07}), i.e., the scalars $Y^A$ as $\textbf{4}_1$ and the fermions $\psi_A$ as $\bar{\textbf{4}}_{-1}$.

Then, since we have a U(1)-neutral $SU(4)$-singlet pseudoscalar mode in the bulk, the issue is whether this $\mathcal{O}_3$ operator is also singlet. Indeed, if we take the matter fields in the original representations, there is a singlet in $ \bar{\textbf{4}} \otimes \textbf{4} \otimes \bar{\textbf{4}} \otimes \textbf{4} $. However, we already argued that the nonsupersymmetric bulk mode agrees to the swapping of the representations $\textbf{s}$ and $\textbf{c}$ of the original ones (\ref{eq18a}) in ABJM. So, the fermions can now sit in $\textbf{8}_s$ while the supersymmetry charges sit in $\textbf{8}_c$. Let us take the singlet spinor field in $\textbf{8}_s = \textbf{1} \oplus \textbf{1} \oplus \textbf{6}$ as one of the $\psi_B$'s, say, $\psi_4 \equiv \psi$, while $Y^A$'s are in the original representation of $\textbf{4}_1$. With these representations, one can simply arrive at a $SU(4) \times U(1)$-singlet from $ \textbf{1} \otimes \textbf{1} \otimes \bar{\textbf{4}} \otimes \textbf{4}$.

Now, by looking at the field equations of the action of (\ref{eq07}), for simplicity and to obtain a right solution, we turn on just one scalar, say, $Y^4 \equiv Y$. Next, we use the following ansatzs
\begin{equation}\label{eq26}
      \psi_{\hat{a}}^a= \frac{\delta_{\hat{a}}^a}{N} \psi, \qquad Y=h(r) I_{N\times N},
\end{equation}
where $h(r)$ is a scalar function on the boundary, $I_{N\times N}$ is the unitary matrix, and the settings for the spinor field are the same as those we already used in \cite{I.N}. So, with the last ansatzs and settings for the matter fields, the potentials of $V_{bos}$ and $V_{ferm}$ vanish. After that, the field equations, for the so-called deformed action, read
\begin{equation}\label{eq27a}
      D_k D^k Y+ \frac{1}{3} tr(\bar{\psi} \psi) Y=0,
\end{equation}
\begin{equation}\label{eq27b}
     i \gamma^k D_k \psi + \frac{1}{3} tr(Y^\dagger Y) \psi=0,
\end{equation}
\begin{equation}\label{eq27c}
    \begin{split}
    & \frac{ik}{4\pi} \varepsilon^{kij} F_{ij}-i\big[Y (D^k Y^{\dagger})-(D^k Y)Y^{\dagger} \big] + \bar{\psi} \gamma^k \psi=0, \\
    & \frac{ik}{4\pi} \varepsilon^{kij} \hat{F}_{ij}- i\big[(D^k Y^{\dagger})Y- Y^{\dagger} (D^k Y) \big] + \bar{\psi} \gamma^k \psi=0,
    \end{split}
\end{equation}
where the $i$ factor appears in front of the Chern-Simons term because of being in the Euclidean space. Next, by taking $Y^{\dagger}=Y$, the second and third terms in both equations of (\ref{eq27c}) are suppressed. \\
On the other hand, we should note that the setting of (\ref{eq26}) is equivalent to considering just the $U(1) \times U(1)$ part of the complete gauge group. Meanwhile, we note that the fundamental matter fields of the ABJM are neutral with respect to the diagonal $U(1)$, which couples to $A_i^+$ of $A_i^\pm\equiv (A_i\pm\hat{A}_i)$, whereas the orthogonal combination $A_i^-$ acts as the baryonic symmetry.\\
Then, from (\ref{eq27c}), we can write
\begin{equation}\label{eq27d}
    \begin{split}
     \frac{i k}{4\pi} \varepsilon^{kij} &F_{ij}^+=-2\bar{\psi} \gamma^k \psi, \\
     &F_{ij}^-=0,
    \end{split}
\end{equation}
and, in addition, to adjust to the bulk, we set $A_i^-=0$. \\
Thereupon, one can simply see that the conditions to satisfy (\ref{eq27a}) and (\ref{eq27b}) together are
\begin{equation}\label{eq28a}
     \partial_k \partial^k h(r)=0, \qquad i \gamma^k \partial_k \psi=0.
\end{equation}
Now, for the scalar and fermion, we use the solution and the ansatz recently applied in \cite{N} and \cite{I.N}, respectively. These are
\begin{equation}\label{eq28b}
    \begin{split}
    & \ \ \ \ \ \ \ \ h = c_7 + \frac{c_8}{r}, \\
    & \psi = \frac{\big(c_9 + i (x-x_0)^k \gamma_k \big)}{\big(c_9^2 + (\vec{u}-\vec{u}_0)^2 \big)^\zeta}\ \eta,
    \end{split}
\end{equation}
where $\eta$ is an arbitrary constant spinor. By putting the $\psi$ ansatz into the relevant equation of (\ref{eq28a}), one can fix its form directly:
\begin{equation}\label{eq28c}
      \psi= \frac{\sqrt{N}}{2} i \sqrt[3/2]{\frac{4}{5}} \frac{(x-x_0)^k \gamma_k}{\big((x-x_0)_k (x-x_0)^k\big)^{3/2}} \bigg(\begin{array}{c}
                                                                                                                  1 \\
                                                                                                                  0
                                                                                                                \end{array}\bigg),
\end{equation}
with a note that $c_9^\dagger =\frac{1}{2} i (x-x_0)_k \gamma^k$ and that, with the Euclidean signature, we have used $\gamma^k=(\sigma_2,\sigma_1,\sigma_3)$ from (\ref{eq11}). \\
Therefore, by using the field equations, the remaining, and of course finite, part of the action and its value read
\begin{equation}\label{eq29}
   S_{modi.}  = - \int_{R^3} d^3r \, (\partial_i h) (\partial^i h), \quad S_{modi.}^{inst.}=- 4\pi c_8,
\end{equation}
where to evaluate the value of the action, we have continued in a similar fashion to \cite{N}, \cite{Gibbons}. In fact, we have used the clear solution of (\ref{eq28b}) with $c_7=1$ and have noted that the contribution from $r=0$ vanishes.

Meanwhile, one can use the solution (\ref{eq28c}) with (\ref{eq27d}) to check that the net magnetic charge of the solution is zero, namely
\begin{equation}\label{eq30}
     B^k=\frac{4\pi i}{3k} \bar{\psi} \gamma^k \psi, \quad \Phi=\oint_s \vec{B}.d\vec{s} =4\pi g=\oint_s F^+=\oint_s \varepsilon^{kij} F_{ij}^+ ds_k =0,
\end{equation}
where $g$ and $\Phi$ show the net magnetic charge and flux, respectively; $\vec{B}$ stands for the magnetic field and $s$ is a round sphere at infinity. This result certifies the $U(1)$ invariance of the boundary solution to be identified with the bulk solution.

Aa a substantial way to verify the dual solutions, we may write the correlation functions of the involved operator in the instanton background. Particularly, the leading contribution for the vacuum expectation value of the operator $\mathcal{O}_3$, in the background, reads
\begin{equation}\label{eq31}
     tr(Y^\dagger Y \bar{\psi} \psi )=\frac{\sqrt{4}}{5\sqrt{5}} \frac{c_8^2}{\big((\vec{u}-\vec{u}_0)^2 \big)^3}.
\end{equation}
This is proportional with $\beta(\vec{u})$ which we gained by analyzing the bulk solution of (\ref{eq16bbcc}) near the boundary. One can also relate the constant coefficients with respect to (\ref{eq24c}). Altogether, we can assert that the boundary and bulk calculations are consistent as expected, while we comment on the other necessary analysis to be done in the next section.

\section{Summary and Further Comments}
In this study, we have found a new instanton solution in the ABJM model as the best sample of AdS$_4$/CFT$_3$ correspondence. We employed an ansatz for the 6-form field strength of ten-dimensional type IIA supergravity while the original background was kept unchanged. After satisfying the field equations and identities, by ignoring a most likely small backreaction on the geometry, we arrived in a fully localized solution in the bulk of Euclidean $AdS_4$. Since the solution appeared because of a D4/M5-brane wrapping around just the associated internal spaces, it was identified with a pseudoscalar. The mode is already known in the spectra of 10d supergravity on $CP^3$ and 11d supergravity on $S^7/Z_k$ when the latter is considered as a U(1) Hopf fibration on the former. It is also notable that for $k=1,2$, the original M2-branes probe the flat space of $R^8$ and $R^8/Z_2$, respectively. For the latter two cases, the situation became somewhat obscure in that at least there was not such a bulk mode in the known spectra of 11d supergravity over $AdS_4 \times S^7/Z_k$. \\
On the other hand, the bulk mode existed when the supercharges were in $\textbf{8}_c$, in contrast to the ABJM supercharges which were in $\textbf{8}_s$. So, to connect the bulk to the boundary, we must switch the representations $\textbf{s}$ and $\textbf{c}$ of ABJM as the resultant theory was then for antimembranes. Next, based on the state-operator correspondence, we found a proper boundary operator of the conformal dimension of $\Delta_+=3$ that responded to the bulk massless pseudoscalar state. The boundary operator was a $SU(4)_R \times U(1)_b$ singlet as the bulk ansatz was so. Afterwards, we observed that to match with the bulk solution, we should just keep a scalar and a fermion next to the $U(1) \times U(1)$ part of the full gauge group. Then, by deforming the action with the proposed operator and solving the resultant boundary equations, while only the mentioned fields were included, we reached a finite action Euclidean solution and saw how the bulk and boundary solutions were mutually compatible. \\
In summary, we can say that we have indeed added a probe anti-D4(M5)-brane to the 3d $\mathcal{N}=6$ $SU(4) \times U(1)$ D2(M2)-brane theory of ABJM that results in a 3d $\mathcal{N}=0$ $SU(4) \times U(1)$ anti-D2(M2)-brane theory, interestingly. The remaining concise discussions are for some other possibly intersecting related points and issues to be carefully addressed later.

The first hint is about supersymmetry. In general, the skew-whiffing procedure breaks all supersymmetries except when the internal space is $S^7$ \cite{NilssonPope}. However, the rigid way to check the ansatzs's supersymmetry is by using the supersymmetry transformations of the gravitinos. By the way, the ansatz of (\ref{eq12}), and also (\ref{eq21a}), obviously breaks all supersymmetries. That is because the branes, to which these fields couple, cover the internal spaces in all generality. In other words, the added branes do not have the right "relative transverse directions" with the branes in the near horizon limit of ABJM to be known as a supersymmetric combination of the branes, according to the well-known brane intersection rules \cite{Bergshoeff}. It is also notable that the magnetic dual of the included D4(M5)-brane is a D2(M2)-brane, some of whose world volume directions are in $AdS_4$. Looking into their behaviors and other related issues may be worthwhile as well.\\
Nevertheless, one may consider some special arrangements of the associated internal spaces on which the branes wind. In other words, one may parametrize, for example, one $CP^1$ with $\theta_1, \varphi_1$ and another $CP^1$ with $\theta_2, \varphi_2$ besides fixing another coordinate, say, $\xi$, to a constant value. Then, the remaining one, $\psi$ in (\ref{eq04b}), may be considered as the coordinate of $S^1$. When the five-dimensional world volume of the added anti-D4-brane wraps around $CP^2 \times S^1$, its effect appears as a point in the lower four-dimensional theory. Its 11d counterpart, according to (\ref{eq21a}), is an anti-M5-brane now wrapping around $CP^2 \times S^1 \times S^1$, where the sixth coordinate is the U(1) fiber coordinate of $\varphi$. For a related typical study look, for instance, at \cite{Chandrasekhar}.

The second hint is about stability. In general, supersymmetry ensures stability. Nevertheless, a nonsupersymmetric solution may be stable in some special cases. In fact, it is justified that all skew-whiffed solutions are stable at least perturbatively \cite{DuffNilssonPope}, and for $S^7/Z_k$, the stability is guaranteed for $k\geq2$. Even so, exactly fixing whether the present solution is stable or not needs a direct study given that the replying instanton actually mediates the converting procedure from the original D2/M2-branes to the skew-whiffed (orientation-reversed) anti-D2/M2-branes. \\
In other words, we should remember that, in ABJM, $k$ D0-branes annihilate into N D4-branes, wrapped around $CP^2 \subset CP^3$, mediated by a NS5-brane instanton. That is, in turn, because of a sort of Higgs mechanism according to which only a $U(1)_b$ is visible in the bulk description \cite{ABJM}, \cite{Bianchi2}. Now, a similar question is what mediates the converting procedure from the original ABJM D2-branes to the skew-whiffed anti-D2-branes? The existing D-instanton is indeed the main reason for doing so. The instanton is practically coming from an anti-D4-brane wrapping around some parts of the internal space of $CP^3$. The resultant effect is breaking all supersymmetries during the process while the conformal transformation is preserved as one can also see from the ansatz structure in (\ref{eq12}). Meanwhile, it is proper here to remember the role of "instantons" as the tunneling agents among various vacua. These vacua are owned by D2-branes and anti-D2-branes for the present case. Still, it is interesting to look at whether one can find other probable "soliton" solutions to accomplish the job of the instanton in breaking all supersymmetries while preserving other symmetries.

From another perspective, we note that a kind of instability may occur because of the probable formation of the brane-antibrane pairs. These pairs, in type II theories, are in general unstable due to the "tachyon" modes on the branes' world volumes \cite{Sen}. As a result, the systems may decay through a process called "tachyon condensation" that is a strong source of instability. Now, the nonsupersymmetric instanton solution here, where an anti-D4-brane is inserted into the background of many D2-branes in the probe approximation, appears to be similar to such a tachyon mode. It is notable that the systems including tachyons generally show the "flip transitions" (from D2-branes to anti-D2-branes here) and that, in $C^4/Z_k$, the flips arise because of the blowups and the blowdowns of the cycles (including the weighted $CP^1$ and $CP^2$) with different dimensions \cite{Narayan}. Therefore, in general, an analysis of the quantum corrections generated by instantons is required to determine whether the squared masses of the fields go up (leading to stable solutions) or go down (leading to unstable solutions) so that some kinds of condensations, which destabilize the vacuum, may occur for the latter.\\
On the other hand, it is shown in \cite{Berkooz} that a kind of instability occurs when there are marginal operators in nonsupersymmetric theories. These operators can destabilize the nonsupersymmetric vacua as soon as $1/N$ corrections could be taken into account. It is further explored in \cite{Murugan} that the "global singlet marginal operators", which appear in the skew-whiffed (orientation-reversed) nonsupersymmetric theories, can disturb the conformal fixed points and may cause instability. Now, for the similar situation with a global singlet marginal operator here, an instability is probable, although a clear and case-by-case analysis is still necessary.

The third hint is about uplifting the gravity solution to the 10- or 11-dimensional parent theories and the issue of backreaction. It is known that the Kaluza-Klein truncations on the fiber and lens spaces, like those in ABJM, are consistent \cite{Gauntlett}. But, for the special fields included, it is a particular study. Handling this issue and uplifting the solution to the higher-dimensional theories as well as trying to estimate the small backreaction will be interesting. \\
We discuss here a little more on the backreaction induced by including the new Euclidean anti-D4/M5-branes on the original ABJM background. As we already mentioned, the energy-momentum tensors for both external and internal spaces have nonzero contributions with respect to the ansatz of (\ref{eq12}) and the solution of (\ref{eq16bbcc}). But, that is a small amount as one can see from the estimated charge of the included branes and from the correction to the main action. Meanwhile, since we are intersected in the behavior of solutions near the boundary and correlation functions for dual operators, we may neglect backreactions. In other words, because, as argued in \cite{Bianchi3} and \cite{Skenderis}, the Einstein and scalar equations decouple next to the boundary, we can study the scalar $f$ equation in a fixed gravitational background that is Euclidean $AdS_4$ here.\\
Nevertheless, it is still important that the instanton solution and its supersymmetric partner are used as sources for the boundary excitations and correct the higher derivative terms in the effective actions. The corresponding studies for $AdS_5$ D-instanton were originally done in \cite{Green}, \cite{Bianchi1}. Equivalently, for the early founded nonperturbative effect in the M2-brane theory, namely the monopole instanton in \cite{Hosomichi}, some instanton corrections were also addressed in \cite{MartinecMcOrist}. Similar calculations would be interesting and in order for the present D/M-instanton too.

The fourth hint is about various probably important and deep studies of the instanton effects on both sides of the duality. Indeed, we should note that the basic estimate of (\ref{eq31}) is the first simple consistency check of the bulk and boundary solutions. Calculations for multipoint and higher-order correlation functions, as well as the D-instanton corrections to the scattering amplitudes of the involved string theories, may be interesting. Next, one can compare the results with the contributions to the Green functions of the composite operators on the field theory side, similar to the AdS$_5$/CFT$_4$ case. One can so survey the instanton solutions carefully and can see how the calculations agree on both sides of the duality. \\
In general, various technical aspects of the studies of the D-brane and world-sheet instantons (for instance, those in \cite{BlumenhagenCveticKachruWeigand} and \cite{Bianchi4}) may be applicable to the current case. For example, due to the presence of instantons, there are fermionic and bosonic zero modes that appear in the phase integral measure. The instanton corrections to the effective actions and zero modes are studied for the case of a monopole instanton in \cite{Hosomichi}, \cite{MartinecMcOrist} and for a "string instanton" in \cite{Sorokin2}. There are also various nonperturbative corrections because of the world-sheet and membrane instantons discussed especially in \cite{Hatsuda} and references therein. It would be interesting to look for similar corrections, do parallel evaluations, and explore the typical relations between the string instanton in \cite{Sorokin2} and the existing D-instanton.

The last issue is about another way to match the bulk instanton to the boundary. Actually, in ten-dimensional type IIB supergravity over $AdS_5 \times S^5$ versus four-dimensional $\mathcal{N}=4$ \ $SU(N)$ Yang-Mills field theory, a similar bulk solution was adjusted with the $SU(2)$ gauge fields on the boundary. We address this issue for the current background in a forthcoming study with respect to some subtle points and differences.

\section{Acknowledgements}
The real science makes people reasonable and sensitive to the surrounding with logical behaviors. However, unfortunately, it seems that some bodies don't know and really understand this. So, it seems that I must thank! the Ilam University officials who are bothering people like me because I am seemingly not at their line. Nevertheless, I would like to thank my mother, father, brothers and sisters in home, who encourage and help me to continue my studies, with a great interest.

\end{document}